\documentclass[aps,prx,twocolumn,superscriptaddress,longbibliography]{revtex4-2}
\usepackage{graphicx,color}
\usepackage{dcolumn}% Align table columns on decimal point
\usepackage{bmpsize}
\usepackage{bbm}% bold math
\usepackage[vcentermath,]{youngtab}
\usepackage{amssymb}
\usepackage{amsmath}
\usepackage{bm}
\usepackage{amsfonts}
\usepackage{epstopdf}
\usepackage[colorlinks, citecolor=blue, linkcolor=blue, urlcolor=blue]{hyperref}
\usepackage{multirow}
\def\s0{\sigma^0}

\def\b0{{\bf 0}}

\begin{document}

\title{Measuring Entanglement without Local Addressing in Quantum Many-Body Simulators via Spiral Quantum State Tomography}
\author{Giacomo Marmorini}
\email{marmorini.giacomo@nihon-u.ac.jp}
\affiliation{Department of Physics, College of Humanities and Sciences, Nihon University, Sakurajosui, Setagaya, Tokyo 156-8550, Japan}
\affiliation{Department of Physical Sciences, Aoyama Gakuin University, Sagamihara, Kanagawa 252-5258, Japan}
\author{Takeshi Fukuhara}
\email{takeshi.fukuhara@riken.jp}
\affiliation{RIKEN Center for Quantum Computing (RQC), Wako, Saitama 351-0198, Japan}
\affiliation{Department of Applied Physics, Graduate School of Advanced Science and Engineering, Waseda University, Okubo, Shinjuku, Tokyo 169-8555, Japan}
\author{Daisuke Yamamoto}
\email{yamamoto.daisuke21@nihon-u.ac.jp}
\affiliation{Department of Physics, College of Humanities and Sciences, Nihon University, Sakurajosui, Setagaya, Tokyo 156-8550, Japan}
\affiliation{RIKEN Center for Quantum Computing (RQC), Wako, Saitama 351-0198, Japan}
\begin{abstract}
Quantum state tomography serves as a key tool for identifying quantum states generated in quantum computers and simulators, typically involving local operations on individual particles or qubits to enable independent measurements. However, this approach requires an exponentially larger number of measurement setups as quantum platforms grow in size, highlighting the necessity of more scalable methods to efficiently perform quantum state estimation. Here, we present a tomography scheme that scales far more efficiently and, remarkably, eliminates the need for local addressing of single constituents before measurements. Inspired by the ``spin-spiral'' structure in magnetic materials, our scheme combines a series of measurement setups, each with different spiraling patterns, with compressed sensing techniques. The results of the numerical simulations demonstrate a high degree of tomographic efficiency and accuracy. Additionally, we show how this method is suitable for the measurement of specific entanglement properties of interesting quantum many-body  states, such as entanglement entropy, under various realistic experimental conditions. This method offers a positive outlook across a wide range of quantum platforms, including those in which precise individual operations are challenging, such as optical lattice systems.
\end{abstract}

\maketitle

\section{Introduction}
Accurate and efficient determination of quantum states is essential for advancing quantum technologies, including digital quantum computers~\cite{Lloyd1996-br,Preskill2018-yj,Bharti2022-vb} and analog quantum simulators~\cite{Georgescu2014-im,Altman2021-te}. Historically, experimental physics research involved testing theoretical models by comparing their predictions with measured values of specific material properties, such as electrical conductivity, magnetic susceptibility, thermal characteristics, and more. In recent years, the evolution of quantum technologies has shifted the focus toward extracting more fundamental information about quantum states and exploring their inherent quantum entanglement.
However, conventional methods for estimating quantum states, typically represented by wavefunctions or more generally by density matrices, require computational resources that grow exponentially with system size~\cite{Paris2004-xq,Hawkes2014-lu}. This challenge has become increasingly urgent as a variety of quantum platforms---such as superconducting circuits~\cite{Devoret2013-bg,Wendin2017-io}, photonic qubits~\cite{Kok2007-ww}, trapped ions~\cite{Blatt2012-mn,Haffner2008-tq}, Rydberg atom arrays~\cite{Adams2020-ut,Browaeys2020-un}, and cold atoms in optical lattices~\cite{Gross2017-am,Schafer2020-hn}---have scaled {up} to unprecedented levels.

Given a density matrix $\rho$ representing a quantum state, the expectation value of any observable $M$ can be computed by taking the trace of the product of $\rho$ and $M$: $\langle M \rangle = {\rm Tr}(\rho M)$. Conversely, the density matrix $\rho$ can be experimentally reconstructed from a set of measurements on various observables that, when taken together, provide sufficient information to uniquely determine the quantum state. This process is generally known as quantum state tomography (QST)~\cite{Paris2004-xq,Hawkes2014-lu}. Given that $\rho$ is a $d \times d$ Hermitian matrix {with trace 1}, it has $d^2 - 1$ independent real parameters. Therefore, to uniquely reconstruct $\rho$, {in principle} one needs a set of $d^2 - 1$  expectation values of {independent} observables: $\langle M_n\rangle$ with $n=1,2,\dots,d^2-1$.

For $N$-qubit systems, where $d=2^N$, the Pauli matrices ($X, Y, Z$) along with the identity $I$ are typically used as a local (single-qubit) basis of Hermitian operators. A  ``tomographically complete'' set of observables can then be constructed 
%from the expectation values of 
as the set of $4^N-1$ operators
\begin{eqnarray}
M_n=\sigma^{i_1}\otimes\sigma^{i_2}\otimes\sigma^{i_3}\otimes\cdots\otimes\sigma^{i_N}/\sqrt{d},\label{Pauli}
\end{eqnarray}
where $i_1i_2\dots i_N$ is the base-4 representation of $n=1,2,\dots,4^N-1$ and $(\sigma^0,\sigma^1,\sigma^2,\sigma^3)=(I,X,Y,Z)$. Thus, the procedure of full tomography requires $3^N$ orthogonal measurement settings, each producing outcomes for  $2^N$  different observables. It then involves averaging the outcomes over many repetitions of the measurements for each setting, upon preparing the target state each time,  to estimate accurate expectation values. The exponential growth of the measurement cost limits full QST to {small systems of several qubits}; for example, more than half a million measurements were required to reconstruct an eight-particle entangled state with trapped ions, achieving a fidelity of $F\approx 0.72$~\cite{Haffner2005-zv}. To mitigate the practical difficulties, various techniques, each specialized in a different way, have been proposed for specific target states, {\it e.g} {those states} that are permutationally invariant~\cite{Toth2010-jx,Schwemmer2014-sp}, {represented} by low-rank density matrices~\cite{Gross2010-mz,gross-11,NIPS2011_e820a45f,Kalev2015-ur,Schwemmer2014-sp}, {described} by matrix product states~\cite{Cramer2010-dx,Lanyon2017-ab}, or uniquely determined by their reduced density matrices~\cite{Xin2017-xw}. {It is worth mentioning a novel approach, namely shadow tomography \cite{aaronson-18,huang-20}, which prescribes how to  extract certain quantities of interest from a quantum state without reconstructing the whole density matrix, using relatively few Pauli measurements; available simulations show great potential, especially for the estimation of linear functions of the density matrix (expectation values of observables) \cite{huang-20}.}

Particularly in the context of quantum simulations of condensed-matter physics and field theory, there is a significant demand for techniques that can reconstruct density matrices that have low rank, but represent a large number of qubits or particles, to study the many-body ground state of certain physically important Hamiltonians. The concept of compressed sensing for recovering a low-rank density matrix from an incomplete set of measurements offers an approach to significantly reduce the number of expectation values required for QST down to $\mathcal{O}(rd \log^2 d)$, where $r$ denotes the rank of $\rho$~\cite{Gross2010-mz}. However, in addition to the issue of the number of measurements, practical difficulties still arise in preparing an enormous number of measurement settings as the system size grows. In a standard implementation of QST, typically based on the Pauli basis~\eqref{Pauli}, one usually needs to rotate local quantization axes into the $X$, $Y$, or $Z$ directions and then read out the qubits for each measurement setting~{\cite{Lanyon2017-zg,Bornet2024-rm}}. This requires precise one-by-one operations and local implementations, which become more complex and costly as the system scales up and integrates more densely.

In this work, we propose an efficient QST protocol based on compressed sensing utilizing a ``spiral" measurement set. A spiral measurement can be achieved by rotating qubits in a manner resembling a spin-spiral configuration observed in chiral magnets~\cite{Kimura2007-gh,Rybakov2016-qw}, with a pitch angle $q$. Since this can be implemented using only the combination of global qubit rotations and the application of a longitudinal ($Z$) field gradient, it drastically reduces the cost of preparing the measurement settings required for QST and is especially beneficial for large-scale quantum systems that may not be completely suited for easy and clean single-qubit (-site) addressing, such as systems of ultra-cold atoms in optical lattices~\cite{Gross2017-am,Schafer2020-hn}. {In fact, in the latter platform the single-site control of atoms is not impossible, but still challenging (especially when compared with trapped ions and atoms in optical tweezers), the fundamental reason being that the distance between qubits (spins) is similar to the size of ``diffraction-limit'' beams (a possible exception is given by spin-flip operations \cite{weitenberg-11}).}

After introducing the protocol for preparing the spiral set of measurements, we demonstrate the efficiency and accuracy of spiral compressed sensing through numerical experiments on a classical computer. As test cases, we consider random pure and low-rank mixed states, for both of which the spiral QST shows a high degree of tomographic efficiency and accuracy. Furthermore, we confirm that the method {yields robust performance against typical experimental noise} when applied to the ground states of several spin Hamiltonians that play a central role in condensed-matter physics. The efficiency of the method is further enhanced by progressively including pitch angles, prioritizing those expected to contribute most significantly to the tomography of the given state. Additionally, we investigate how spiral QST captures specific entanglement properties in intriguing many-body quantum states. This demonstrates its potential to advance quantum simulation studies as a key technique for exploring fundamental quantum phenomena from the perspective of quantum information science.

\section{Quantum state tomography via spiral compressed sensing}
\label{sec2}
\subsection{Methodology}

\begin{figure}[tb]
\begin{center}
\includegraphics[scale=0.46]{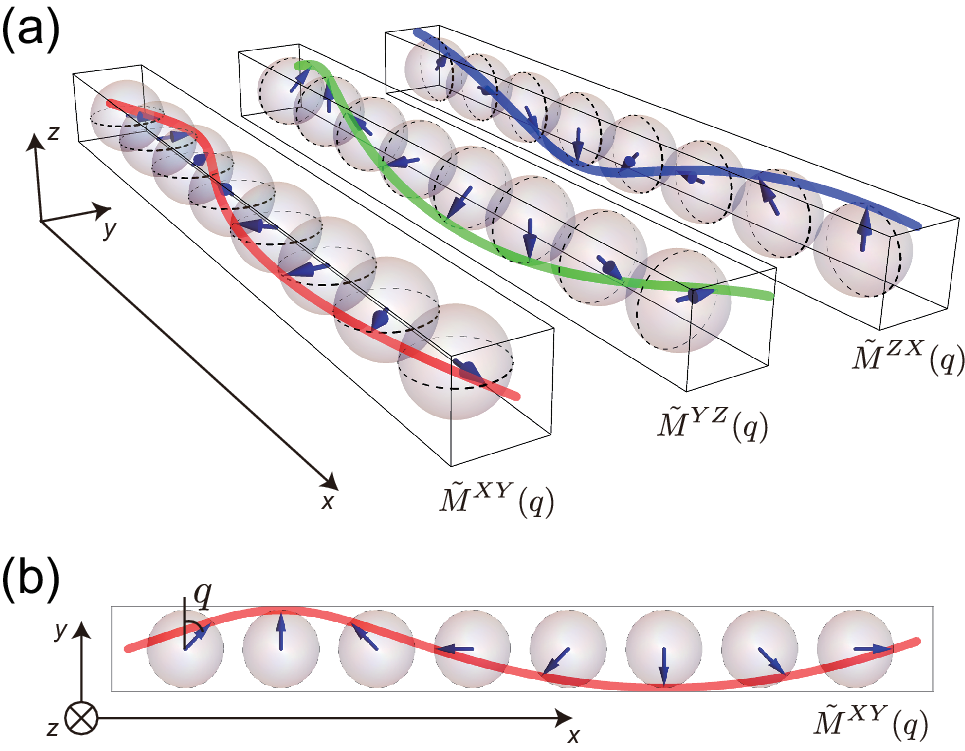}
\end{center}
\caption{{Spiral measurements.} (a) Illustrations of the three spiral measurement {sets}. The blue arrows indicate the local measurement axis at each qubit. (b) Pitch angle $q$ in the spiral plane of $\tilde{M}^{XY}(q)$.\label{fig_spiral}}
\label{effrate}
\end{figure}

We first introduce the protocol for QST via compressed sensing using a set of spiral measurement operators (Fig.~\ref{fig_spiral}{(a)}). Conventional compressed-sensing QST~\cite{Gross2010-mz,gross-11,NIPS2011_e820a45f,Kalev2015-ur} reconstructs a pure or nearly-pure quantum state, represented by a low-rank density matrix, using $\mathcal{O}(rd \log^2 d)$ random samples from the Pauli measurements~\eqref{Pauli}. The idea of the spiral compressed sensing is to use measurements based on three sets of spiral-spin operators, instead of Pauli operators: 
\begin{equation}
\begin{split}
\tilde{M}^{XY}(q)&=Z^{(1)\prime}\otimes Z^{(2)\prime}\otimes Z^{(3)\prime}\otimes\cdots\otimes Z^{(N)\prime}/\sqrt{d},\\
\tilde{M}^{YZ}(q)&=Z^{(1)\prime\prime}\otimes Z^{(2){\prime\prime}}\otimes Z^{(3){\prime\prime}}\otimes\cdots\otimes Z^{(N){\prime\prime}}/\sqrt{d},\\
\tilde{M}^{ZX}(q)&=Z^{(1){\prime\prime\prime}}\otimes Z^{(2){\prime\prime\prime}}\otimes Z^{(3){\prime\prime\prime}}\otimes\cdots\otimes Z^{(N){\prime\prime\prime}}/\sqrt{d}, \label{spiral1}
\end{split}
\end{equation}
with
\begin{equation*}
\begin{split}
Z^{(i)\prime}&=\cos (q (i-i_0))X+\sin (q (i-i_0))Y,\\
Z^{(i){\prime\prime}}&=\cos (q (i-i_0))Y+\sin (q (i-i_0))Z,\\
Z^{(i){\prime\prime\prime}}&=\cos (q (i-i_0))Z+\sin (q (i-i_0))X, \label{spiral2}
\end{split}
\end{equation*}
where $N$ qubits (or sites) are supposed to be arranged along the $x$-axis at equal intervals, indexed by $i=1,2,\dots,N$. Here, $q$ and $i_0$ represent the pitch and origin of the spiral structure of the measurement axis [see Fig.~\ref{fig_spiral}{(b)}]. Measuring a spiral operator, say $\tilde{M}^{XY}(q)$, on a given quantum state, or equivalently measuring $Z^{(i)\prime}$ on each qubit, can be carried out as follows: First, $Z$-axis rotation gates $R_Z (\theta_i)$ are applied to every qubit with linearly varying angles with pitch $q$: $\theta_i =-q (i-i_0)$. Subsequently, the $X$-basis measurements are performed for all the qubits by applying the Hadamard gates to them, followed by measurements in the computational basis ($Z$-basis) in a standard way. The other spirals, $\tilde{M}^{YZ}(q)$ and $\tilde{M}^{ZX}(q)$, can also be measured similarly by applying additional uniform gates to change the spiral plane into the $xy$-plane before the $R_Z (\theta_i)$ rotations.

One of the advantages of this procedure is that it can be implemented without the need for complicated single-site addressing. For example, in systems like trapped ions or cold atoms, where the qubit states are typically defined by the magnetic sublevels of an atom or ion, the spiral rotation around the $Z$-axis can be achieved by applying a magnetic-field gradient whose strength varies linearly as a function of the site index $i$, with $i_0$ as the origin:
\begin{equation}
H_{\rm grad}= -{\frac{B}{2}} \sum_i (i-i_0) Z_i, \label{Hgrad}
\end{equation}
where $O_i$ denotes the  {single-qubit} operator $O$ acting on site $i$, for a short time interval of $\Delta t=q/B$. The spiral measurement setups with three different planes [shown in Fig.~\ref{fig_spiral}{(a)}] can then be prepared by combining this with globally uniform $\pi$/2 Rabi pulses~\cite{Fukuhara2015-sn,Mazurenko2017-yb,Brown2017-fp}. This process can also be viewed as the inverse of creating a spin-spiral state, starting from the initial ``all-up'' state~\cite{Koschorreck2013-ur,Bardon2014-bs,Hild2014-mz,Jepsen2020-ze,Jepsen2021-xl,jepsen-22}.

By tuning the pitch $q$ through {the manipulation of} the strength $B$ and/or the application time $\Delta t$ of the field gradient, different measurement setups can be prepared for each spiral plane. When denoting the number of different pitch angles by $n_q$, the number of independent experimental setups is $3n_q$, where the factor of 3 comes from the three spiral planes. For each setup, $2^N-1$ mutually independent expectation values can be obtained, as it is possible to choose whether or not to include the measurement of each individual qubit in the calculation of the multi-site correlation functions. In other words, from a single setup, say $\tilde{M}^{XY}(q)$ with a specific $q$, one can simultaneously obtain $\langle \tilde{M}^{XY}(q) \rangle$ and all the correlation functions where any of its $Z^\prime_i$s are replaced with $I$. We denote the corresponding operators by $\tilde{M}^{\alpha\beta}_{\bm{a}}(q)$, where $\alpha\beta=XY,YZ,ZX$ and $\bm{a}\in \{0,1\}^N$ is a multi-index indicating the positions of the identity matrices (labeled by 0), {\it e.g.}, $\tilde{M}^{XY}_{(1,0,\ldots,0)}(q)= Z^{(1)\prime}\otimes I \cdots \otimes I/\sqrt{d}$, $\tilde{M}^{YZ}_{(1,1,0,\ldots,0)}(q)= Z^{(1)\prime\prime}\otimes Z^{(2)\prime\prime}\otimes I \cdots \otimes I/\sqrt{d}$, $\tilde{M}^{ZX}_{(1,\ldots,1)}(q)= Z^{(1)\prime\prime\prime}\otimes\cdots \otimes  Z^{(N)\prime\prime\prime}/\sqrt{d}=\tilde{M}^{ZX}(q)$, etc. Note that this family includes the product of only identities at all sites, which contains no information. Consequently, the number of {independent} measurements that can be obtained from the three spiral sets with $n_q$ different pitch angles is, in total, $\sim 3n_q 2^N$, which can be made comparable to $\mathcal{O}(rN^22^N)$, the required number of Pauli measurements for successful compressed sensing~\cite{Gross2010-mz}, by choosing a sufficiently large $n_q$.

Here, it should be noted that the spiral operators $\{\tilde{M}^{\alpha\beta}_{\bm{a}}(q)\}$ present a fraction of non-orthogonal pairs and may contain some linear dependencies (for some choices of $q$'s), unlike the Pauli basis~\eqref{Pauli}. Additionally, the spiral basis is ``structured,'' whereas the random choice from Pauli measurements is not. Thus, it is not trivial to establish that compressed sensing will also work effectively with spiral measurements, as will be demonstrated below. 

Although the examples provided here consider qubit arrays arranged in a straight line, it is straightforward to extend this approach to two or more dimensions by employing a pitch vector $\bm{q}$.

\subsection{Numerical tests} \label{sec:numtest}
To demonstrate the efficiency and accuracy of the spiral QST, we test the reconstruction of random pure states ($r=1$) and rank-3 mixed states ($r=3$). For the pitch angle $q$, we employ {a set of $2N$} values with intervals of $\pi/N$, defined as
\begin{eqnarray}
q=\frac{\pi}{N}l~~(l=-N+1,-N+2,\dots,N),\label{pitch}
\end{eqnarray}
and the origin (zero point) of the field gradient is set to the center of the chain: \begin{eqnarray}
i_0=\left\{\begin{array}{cl}
    N/2&   \text{for odd } N,\\
    (N+1)/2 &   \text{for even } N.
\end{array}\label{origin}
\right.
\end{eqnarray}
%but introducing an offset to this origin does not affect the overall qualitative results. 
The above parameters, while reasonable from a physical standpoint, are by no means the only possible choice.
We utilize the singular value thresholding (SVT) {algorithm}~\cite{Gross2010-mz,Cai2008-zn} for compressed sensing, sampling expectation values of spiral operators in the set $\{\tilde{M}^{\alpha\beta}_{\bm{a}}(q)\}$
%from Eqs.~\eqref{spiral1} 
associated to {$n_q$} randomly chosen pitch angles from Eq.~\eqref{pitch} (see Appendix~\ref{appendixA}). Target states $\rho$ are generated by randomly selecting from the Haar measure on a $d\times r$-dimensional system ($d=2^N$) and tracing out the $r$-dimensional ancilla. To model typical noise during the preparation of the target state, we add depolarizing noise of strength $\gamma$ after generating the target states as $\rho\rightarrow (1-\gamma)\rho +\gamma I^{\otimes N}/d$. We also account for measurement noise modeled by Gaussian fluctuations with standard deviation $\sigma$ in calculating expectation values for each measurement: $\langle \tilde{M}^{\alpha\beta}_{\bm{a}}(q) \rangle\rightarrow \mathcal{N}(\mu=\langle \tilde{M}^{\alpha\beta}_{\bm{a}}(q) \rangle,\sigma)$

\begin{figure}[tb]
\begin{center}
\includegraphics[scale=0.42]{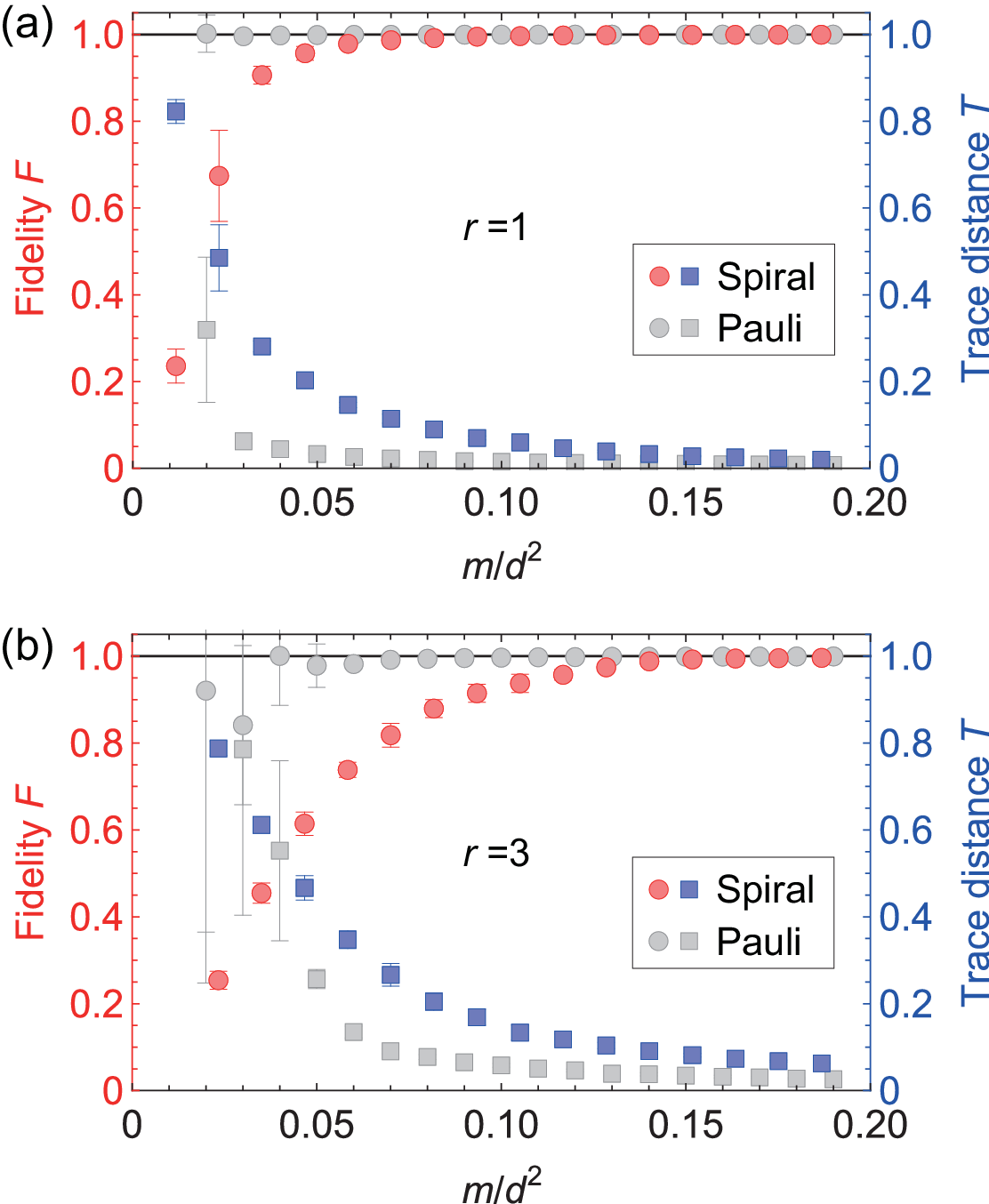}
\end{center}
\caption{{Numerical tests of spiral QST.} Sample-averaged fidelity and trace distance for compressed-sensing tomography of (a) random pure states (rank $r=1$) and (b) random mixed states (rank $r=3$) of 8 qubits, plotted as functions of the number of {independent} measurements $m$ scaled by $d^2=4^8$. The sampled states include 5 percent depolarizing noise ($\gamma=0.05$) and measurement Gaussian statistical noise with a standard deviation of $\sigma=0.1/d$. Spiral and Pauli measurements are compared in each case. For $r=3$, the results of Pauli-based tomography {under the same conditions} were also reported in previous work~\cite{Gross2010-mz}.} \label{fig_random}
\end{figure}

For evaluating the accuracy of the state reconstruction, we use the fidelity 
\begin{eqnarray*}
F(\rho_1,\rho_2)={\rm Tr} \left(\sqrt{\sqrt{\rho_1}\rho_{2}\sqrt{\rho_1}}\right)^2\Big/{\rm Tr} (\rho_1){\rm Tr} (\rho_2)
\end{eqnarray*}
and the trace distance
\begin{eqnarray*}
T(\rho_1,\rho_2)=||\rho_1-\rho_2||_1/2,
\end{eqnarray*}
where $||A||_1\equiv {\rm Tr}(\sqrt{A^\dagger A})$ {is the trace norm (or nuclear norm)}. These metrics take values 1 and 0, respectively, if and only if the two states $\rho_1$ and $\rho_2$ are identical. In Figs.~\ref{fig_random}{(a)} and~\ref{fig_random}{(b)}, we present $F(\rho,\rho^{\rm rec})$ and $T(\rho,\rho^{\rm rec})$ between the target state $\rho$ and the reconstructed state $\rho^{\rm rec}$ for pure and mixed states, respectively, of $N=8$ qubits, plotted as functions of the number of spiral measurements (expectation values), $m$, used in the SVT algorithm, expressed in units of $d^2$. The results include noise levels corresponding to {$\gamma=0.05$} and $\sigma=0.1/d$, matching those used in the demonstrations with Pauli-basis measurements in Ref.~\cite{Gross2010-mz}. 

As seen from the results for $r=1$, the performance of spiral QST is comparable to (though slightly worse than) the conventional Pauli-basis compressed sensing~\cite{Gross2010-mz}. The fidelity and trace distance indicate that the state reconstruction is very successful with only about 10 percent of a tomographically complete set of measurements. For mixed states with rank $r=3$, the results are also notable. %(see Ref.~\cite{Gross2010-mz} for comparison with the Pauli-basis case). 
Given that spiral measurements do not require precise single-site addressing nor a large number of experimental setups, this outcome highlights the practicality of our approach. Building on these results for random quantum states, we now aim to explore more realistic scenarios reflecting actual experimental conditions in the remainder of the paper.

{Before that, let us note that one might be interested in comparing $\rho^{\rm rec}$ not with the ideal reference state $\rho$, but with the actual experimentally prepared state, which, in real situations, is affected by many sources of noise and therefore unknown. In this case one can make use of the ``minimax'' method for fidelity estimation developed in Refs.~\cite{seshadri-24l,seshadri-24a}. Though it assumes a pure reference state, it does not impose any restrictions on the measurement operators (such as random drawing from a basis of Hermitian operators \cite{flammia-11}), and it is possible to simply employ  some of the spiral operators that were not used for the tomographic reconstruction; also, the sample complexity is comparable to that of the renowned ``direct fidelity estimation'' method \cite{flammia-11}. Even if $\rho^{\rm rec}$ is mixed, we can still try to find a pure state that is close to it in fidelity to employ in the minimax method and then use the triangle inequality to estimate the fidelity between $\rho^{\rm rec}$ and the experimentally prepared state. }

\section{Simulation of real-world experiments}
\label{sec:realworld}

\subsection{Spiral-QST for typical ground states} \label{sec:simexp}
Quantum state reconstruction for large-scale systems is especially critical in quantum simulations of, {\it e.g.}, condensed-matter physics and quantum field theory, where understanding the behavior of the ground state as it approaches the thermodynamic or continuum limit is essential. In this subsection, we focus on simulating the spiral QST for a more specific experimental scenario. The target state will be the ground state of the prototypical spin-1/2 Heisenberg chain with open boundary conditions, described by the Hamiltonian
\begin{eqnarray}
\mathcal{H}=\sum_{i=1}^{N-{p}}\sum_p J_p\left(X_iX_{i+p}+Y_iY_{i+p}+Z_iZ_{i+p}\right),\label{Heisenberg}
\end{eqnarray}
where $J_p$ represents the spin-exchange coupling between the $p$-th nearest neighbors.

\begin{figure}[tb]
\begin{center}
\includegraphics[scale=0.45]{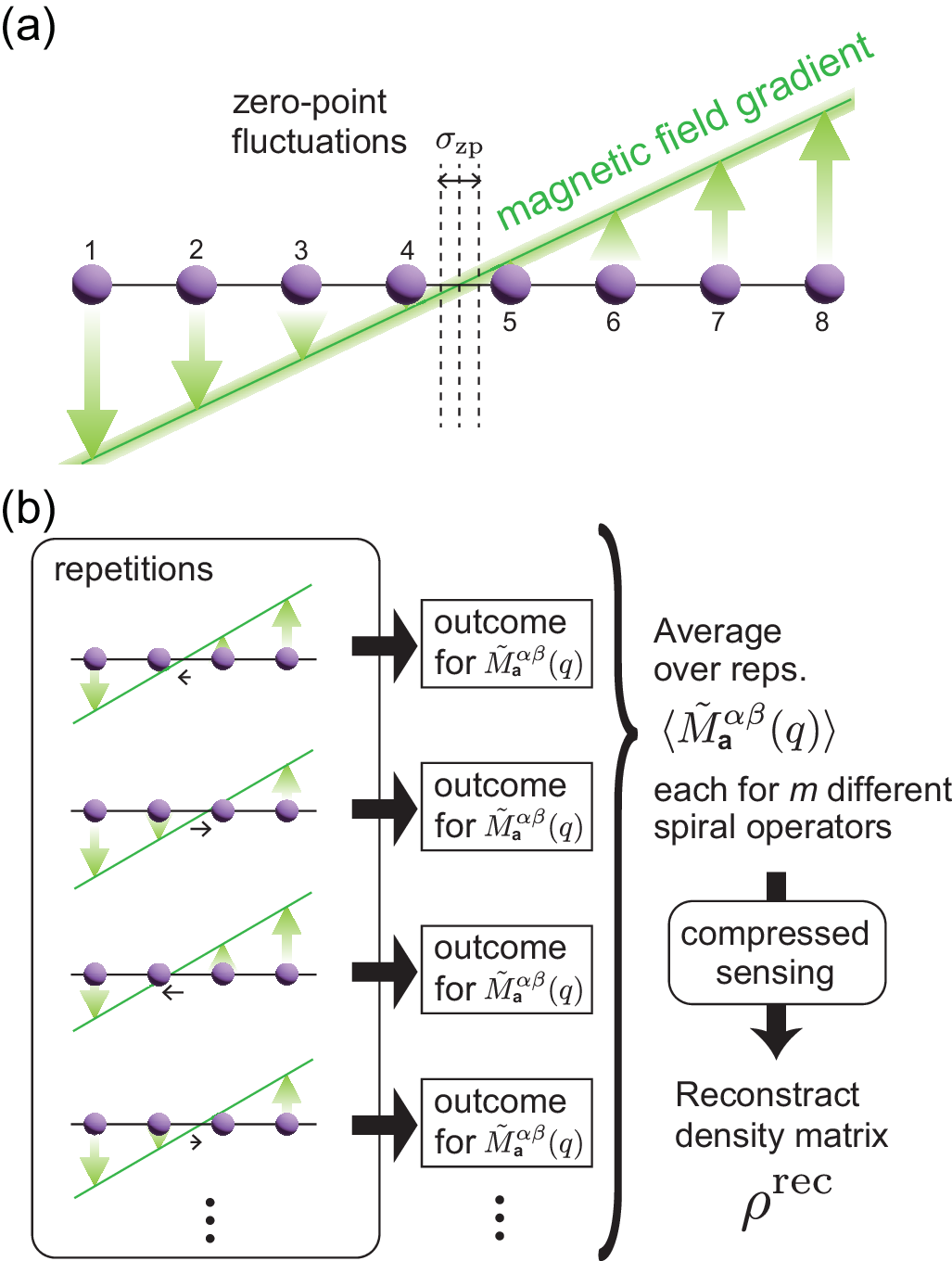}
\end{center}
\caption{{Magnetic field gradient method for spiral QST.} (a) Magnetic field gradient used to create spiral-spin settings. Typical experimental noise arises from fluctuations in the strength of the magnetic field over time. These fluctuations cause variations in the zero point of the gradient, modeled as Gaussian fluctuations with a mean of 0 and a standard deviation $\sigma_{\rm zp}$ (in units of lattice spacing). (b) Protocol for numerical simulations of spiral QST with magnetic field gradient. We simulate a realistic experimental process to measure observables by averaging the outcomes over many shots of repeatedly prepared target states for each experimental setting of spiral plane $\alpha\beta=XY$, $YZ$, or $ZX$ and pitch $q$. Each repetition is subject to fluctuations in the magnetic field gradient, introducing noise into the expectation values of the spiral operators.}
\label{fig_zeropoint}
\end{figure}

The spin-1/2 Heisenberg model, a foundational model for studying quantum magnetism, has been realized in the strong interaction regime of systems consisting of two-component ultracold Fermi~\cite{Boll2016-wd,Mazurenko2017-yb,Brown2017-fp} or Bose~\cite{Fukuhara2015-sn,Jepsen2020-ze,Jepsen2021-xl,Sun2021-mv} atoms trapped in an optical lattice. {Antiferromagnetic correlations} have been observed in both one~\cite{Boll2016-wd} and two~\cite{Cheuk2016-mw,Mazurenko2017-yb,Brown2017-fp} dimensions using spin-selective imaging with a quantum-gas microscope (QGM)~\cite{Gross2021-iv}, which enables site-resolved longitudinal ($Z$-basis) measurements. {The correlation length of the observed antiferromagnetic correlations extends to  approximately 10 sites in Ref.~\cite{Mazurenko2017-yb}}. Transverse ($X$- or $Y$-basis) measurements are also possible by applying a global $\pi/2$ rotation before imaging~\cite{Mazurenko2017-yb,Brown2017-fp}. In these typical optical-lattice systems, however, the complete reconstruction of density matrices, which conventionally requires site-resolved rotations of spin directions, has not yet been achieved, partly due to the short lattice spacing imposed by the laser wavelength used to form the optical lattice.

Below, we numerically simulate the spiral compressed-sensing QST for Heisenberg spin systems, assuming that spiral measurements are implemented through the combined application of a magnetic-field gradient $H_{\rm grad}$ as in Eq.\eqref{Hgrad}~\cite{Hild2014-mz,Jepsen2020-ze,Jepsen2021-xl,jepsen-22} and global $\pi/2$ Rabi pulses~\cite{Fukuhara2015-sn,Mazurenko2017-yb,Brown2017-fp}, followed by $Z$-basis QGM measurements. To account for potential experimental errors beyond the abstract noise models parameterized by $\gamma$ and $\sigma$ in the previous subsection, we consider more concrete scenarios. In realistic experimental settings, a significant source of measurement noise could arise from fluctuations in the zero point of the magnetic field gradient, modeled as $-B (i-i_0) \rightarrow -B (i -i_0- \delta_{\rm zp})$, where $\delta_{\rm zp}$ represents the zero-point displacement in units of the lattice spacing [see Fig.~\ref{fig_zeropoint}{(a)}]. This displacement leads to an unwanted phase shift in the spiral structures, causing the experimentally obtained values to correspond to spiral operators with $q(i-i_0)$~[Eq.~\eqref{spiral1}] replaced by $q(i -i_0-\delta_{\rm zp})$. We assume that $\delta_{\rm zp}$ is Gaussian distributed with a mean of 0 and a standard deviation $\sigma_{\rm zp}$. It should also be noted that {the accurate determination of the} expectation values requires a sufficiently large number of repetitions of the same measurements for each spiral setting. For each experimental setup --- defined by a specific spin-spiral plane and pitch angle $q$ --- we simulate multiple QGM shots, each affected by the zero-point fluctuations of the magnetic field gradient as described above [see Fig.~\ref{fig_zeropoint}{(b)}]. An adequate number of repetitions is necessary to capture the quantum nature of the state, {namely} the quantum fluctuations inherent in measuring each spiral operator.

First, we consider the test case of the ground state $|\Psi_0\rangle$ of the $N=8$ Heisenberg model~\eqref{Heisenberg} with only nearest-neighbor (NN) antiferromagnetic coupling ($J_1=1$ and $J_{p\neq 1}=0$), which can easily be computed using the exact diagonalization method. {For highly symmetric states, such as the ground state of typical condensed-matter Hamiltonians, the majority of the QST weight is expected to concentrate on spiral operators associated with a limited number of pitch angles; in addition, the pitch angles are related by the Hamiltonian's symmetry (which is not spontaneously broken at finite size).
This observation suggests a natural strategy: constructing an expansion that prioritizes the most relevant pitch angles as well as the symmetries, and planning the experiment based on this hierarchy. Accordingly, we perform spiral QST by progressively including pitch angles from the set defined in Eq.~\eqref{pitch}, starting with $\{l=0\}$, followed by $\{l=0,1, -1\}$, $\{l=0,1,-1,2,-2\}$, and so on (see Appendix~\ref{appendixB} for an analysis of relevant pitch angles). Note that all these sets respect inversion symmetry.}
% For highly symmetric states, such as the ground state of typical condensed-matter Hamiltonians, the majority of the QST weight is expected to concentrate on spiral operators associated with a limited number of pitch angles. This observation suggests a natural strategy: constructing an expansion that prioritizes the most relevant pitch angles and planning the experiment based on this hierarchy. Accordingly, we perform spiral QST by progressively including pitch angles from the set defined in Eq.~\eqref{pitch}, starting with ${l=0}$, followed by ${l=0,1}$, ${l=0,1,2}$, and so on (see Appendix~\ref{appendixB} for an analysis of relevant pitch angles). 

\begin{figure}[tb]
\begin{center}
\includegraphics[scale=0.42]{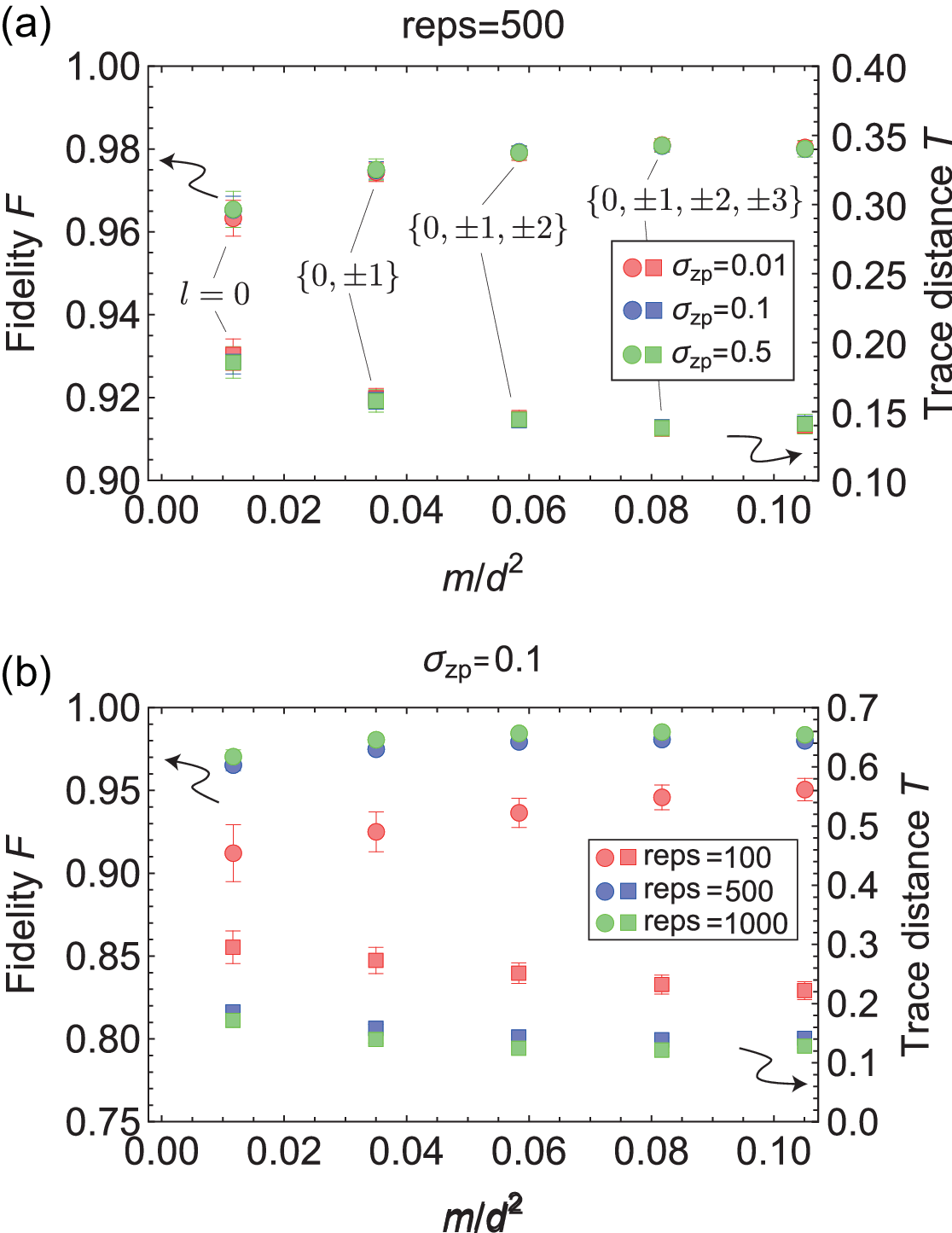}
\end{center}
\caption{{Spiral QST for the ground state of the NN Heisenberg chain.} Sample-averaged fidelity and trace distance for spiral compressed-sensing QST of the ground state of the 8-site Heisenberg chain with NN antiferromagnetic spin-exchange coupling, plotted as functions of the number of measurements $m$ scaled by $d^2=4^8$. In ({a}), the results for different amplitudes of magnetic field fluctuations {($\sigma_{\rm zp}=0.01,0.1,0.5$)} are compared, with the number of repetitions for evaluating each expectation value fixed at {${\rm reps}=500$}. {The dependence on $\sigma_{\rm zp}$ is essentially absent, which is supported by the symmetry argument given in the main text.} In ({b}), the results for different numbers of repetitions (${\rm reps}=100, 500, 1000$) are compared {(the amplitude of magnetic field fluctuations is chosen as $\sigma_{\rm zp}=0.1$ for concreteness)}. We adopt the strategy of progressively including expected relevant pitch angles {in symmetric fashion; the data points with the fewest measurements correspond to the spiral QST with $\{l=0\}$, followed by $\{l = 0, 1,-1\}$, $\{l = 0, 1,-1, 2,-2\}$, and so on.} }
\label{fig_Heisenberg}
\end{figure}

In Fig.~\ref{fig_Heisenberg}{(a)}, we present the fidelity and the trace distance between the target state $\rho=|\Psi_0\rangle\langle\Psi_0|$ and the reconstructed state $\rho^{\rm rec}$ obtained via the spiral QST, with the number of repetitions (denoted by ``reps'') fixed at {${\rm reps}=500$} for evaluating each expectation value. We compare the results for different amplitudes of magnetic field fluctuations ({$\sigma_{\rm zp}=0.01,0.1,0.5$}); other sources of noise are absent. First, we observe that the dependence on $\sigma_{\rm zp}$ {appears to be absent,} {except for variations attributable only to statistical errors}. Second, our ``relevant-pitch-angles'' strategy appears to be highly effective. The fidelity reaches approximately $0.98$ for $m/d^2\approx 0.1$, which corresponds to only 20-30 independent measurement setups, considering that one can obtain $d=2^N$ independent correlation functions from each experimental setup. The data points with the fewest measurements correspond to the results obtained by using just $m=3\times2^8-2$ expectation values, based  merely on three experimental setups of uniform $X$-, $Y$-, and $Z$-measurements (i.e., $l=0$ or $q=0$) in the SVT method. This condition already achieves very good fidelity and trace distance, likely due to the simplicity of the ground state of the NN Heisenberg model. 
% Interestingly, the fidelity decreases (and the trace distance increases) when additional measurements are included in the small $m/d^2$ range; this may simply be due to the possibility that a better optimization of the SVT algorithm for $l=0$ was achieved in our attempts, but further clarifications would be worthwhile. 

The absence of dependence on the zero-point fluctuations of the field gradient can be understood from symmetry arguments. From Eq.~\eqref{spiral1}, a shift in the zero point, $\delta_{\rm zp}$, affects the spiral measurement, say in the $XY$ plane, at site $i$ as a unitary rotation around the $Z$-axis by angle $q\delta_{\rm zp}$ (independent of $i$) as follows:
\begin{eqnarray*}
&&\cos (q (i-i_0-\delta_{\rm zp}))X+\sin (q (i-i_0-\delta_{\rm zp}))Y\\
&=&e^{i q\delta_{\rm zp} Z/2} Z^{(i)^\prime} e^{-i q\delta_{\rm zp} Z/2}.
\end{eqnarray*}
Thus, if the state of the system is symmetric under global rotations around the $Z$-axis, the expectation value $\langle \tilde{M}^{XY}_{\bm{a}}(q)\rangle$ remains entirely unaffected by the shift $\delta_{\rm zp}$. Moreover, the total set of possible outcomes (the spectrum of $\tilde{M}^{XY}_{\bm{a}}(q)$) remains unchanged under the unitary rotation in the present case, and the Born rule ensures that the probability of each measurement outcome is invariant. Therefore, spiral measurements in the $XY$ plane are inherently robust against zero-point fluctuations of the field gradient for the state invariant under global rotations around the $Z$-axis, even when $\delta_{\rm zp}$ varies between repetitions as illustrated in Fig.~\ref{fig_zeropoint}. A similar argument applies to spiral measurements in the $YZ$ and $ZX$ planes. Since the target state currently considered exhibits symmetry under global rotations around all $X$, $Y$, and $Z$ axes due to the SU(2) spin symmetry of the Heisenberg model, the spiral QST results are perfectly robust against noise arising from any zero-point shifts in the field gradient, not limited to Gaussian-type fluctuations.

In Fig.~\ref{fig_Heisenberg}{(b)}, we compare the results obtained by varying the number of repetitions for each experimental setting (i.e., the number of QGM shots after each unitary transformation corresponding to spiral operators defined by a specific pitch angle $q$ and spin plane) while keeping the amplitude of magnetic field fluctuations fixed at $\sigma_{\rm zp}=0.1$, {although a different value would not change the results, as discussed above.} We observe that this dependence is significant, at least below a certain threshold, although this sensitivity is a general issue for QST. Our results suggest that ${\rm reps} \gtrsim 500$ is required to achieve convergence.

\begin{figure}[tb]
\begin{center}
\includegraphics[scale=0.42]{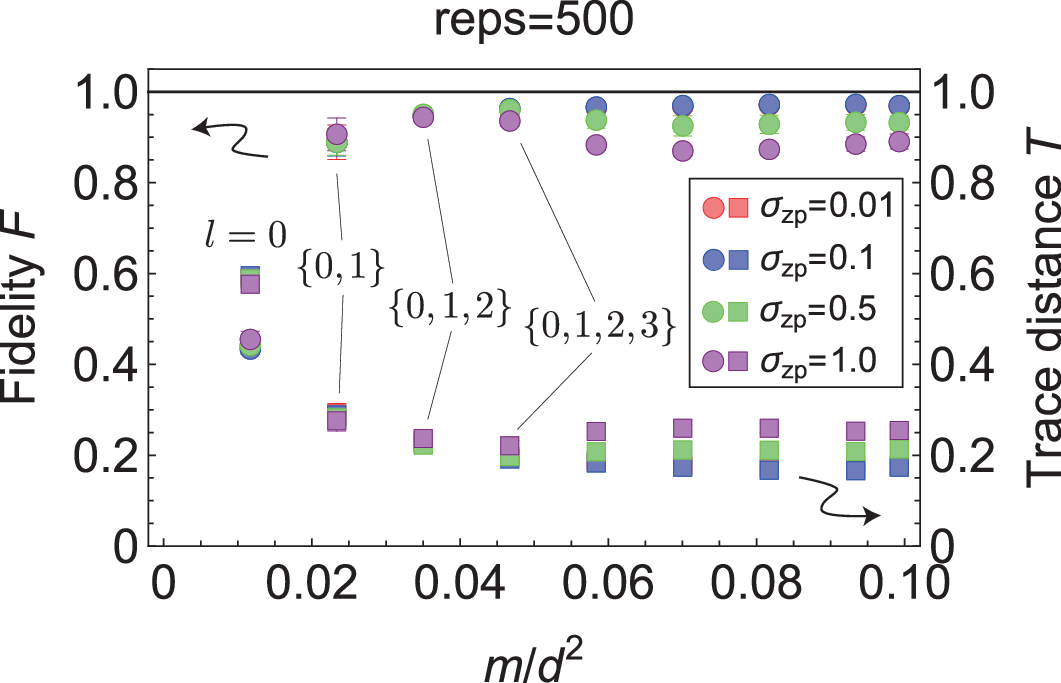}
\end{center}
\caption{{Spiral QST for the ground state of the NN Heisenberg + DM chain.} Sample-averaged fidelity and trace distance for spiral compressed-sensing QST of the ground state of the {8-site} Heisenberg chain with NN ferromagnetic spin-exchange coupling and DM interactions of equal strength, plotted as functions of the number of measurements $m$ scaled by $d^2=4^8$. {The results for different amplitudes of magnetic field fluctuations ($\sigma_{\rm zp}=0.01,0.1,0.5,1.0$) are compared, with the number of repetitions for evaluating each expectation value fixed at ${\rm reps}=500$; note that the data for $\sigma_{\rm zp}=0.01$ and $\sigma_{\rm zp}=0.1$ are nearly indistinguishable.} The data points with the fewest measurements correspond to the spiral QST with $l=0$, followed by $l = 0, 1$, $l = 0, 1, 2$, and so on. }
\label{fig_DM}
\end{figure}

We have so far presented the test case of the NN Heisenberg model. As shown above, in this case, the uniform measurements ($l=0$ or $q=0$) seem to provide the major contributions to the QST of the ground state. However, this is not always the case. To demonstrate this, we now discuss another simple example where the so-called Dzyaloshinskii-Moriya (DM) interaction 
\begin{eqnarray}
-D\sum_{i=1}^{N-1}\left(X_iY_{i+1}-Y_iX_{i+1}\right),\label{DMint}
\end{eqnarray}
exists in addition to the NN Heisenberg coupling. The DM interactions are known to induce non-collinear spin structures, expected to change the most relevant pitch angles in the spiral QST. {In addition, the DM interactions reduce the ground state's symmetry from SU(2) symmetry to U(1), corresponding to global rotations around the $Z$-axis only. This raises the question of robustness of the method against varying levels of noise caused by magnetic field fluctuations when the target state exhibits only a subset of spin-rotational symmetry.}

In Fig.~\ref{fig_DM}, we present the results for the ground state of the {$N=8$} Heisenberg + DM chain with $D=-J_1=1$ and $J_{p\neq 1}=0$, {adopting the pitch-angle sequence $\{l=0\}$, $\{l=0,1\}$, $\{l=0,1,2\}$, etc.}
%\textcolor{green}{adopting the same sequence of pitch angles as in the previous case without DM interactions.} 
As seen in the figure, uniform measurements with $l=0$ alone are insufficient for accurately reconstructing the ground state in this case. The DM interaction introduces additional complexities, requiring the inclusion of finite pitch angles to capture the non-collinear spin correlations present in the ground state. Specifically, we identified the most relevant pitch angle (taken individually) in this case as {$l=4$} or $q=\pi/2$ (see Appendix~\ref{appendixB}).

% ??\textcolor{green}{ For larger-size system realized in experiments, convergence might be achieved with fewer measurements by employing a strategy that prioritizes the most relevant pitch angles, guided by simulations on smaller systems. Nevertheless, as demonstrated in Fig.~\ref{fig_DM}, the compressed sensing with the SVT method already performs remarkably well, even when the most relevant pitch angle ($l=4$ in this case) is not explicitly included.}

As expected from the absence of the global SU(2) symmetry, an appreciable dependence on $\sigma_{\rm zp}$ is indeed present, although the results converge very well around and below $\sigma_{\rm zp} = 0.1$. The fluctuations with $\sigma_{\rm zp}=0.01$ roughly correspond to an experimental accuracy of 10 {\textmu}Gauss on the magnetic field {\footnote{{If one considers a typical lattice spacing of roughly $500$~nm and a typical magnetic field gradient of $20$ G/cm (the different $q$'s can then be obtained by varying the application time), then the variation of the field over $0.01 \times$ (lattice spacing) is about $10$ $\mu$Gauss. In other words, if the fluctuations at a point in space are bigger than $10$ $\mu$Gauss, the shift in the zero-point in general will be bigger than $0.01 \times$ (lattice spacing).}}}, 
which, while attainable, remains challenging {\cite{xu-19}}. These results suggest that such extremely precise control is not necessary for our protocol, at least under the assumption of a symmetric zero-point distribution (such as a Gaussian).
{For $\sigma_{zp}>0.1$, after the data essentially reach the saturation of fidelity (for the given conditions) around $m/d^2\sim 0.04$, the tradeoff between adding more measurements and their additional noise  becomes unfavorable (also, for high $q$ the effect of $\sigma_{zp}$ is higher). One can find different sequences of $q$ sets for which the monotonicity is preserved up to at least $\sigma_{zp}=0.5$, but at the same time the tomographic performance is generally poorer (see also the discussion at the end of Appendix~\ref{appendixB}).}
 
\subsection{Detection of entanglement measures} \label{sec:detee}
One of the primary purposes of reconstructing full density matrices, rather than merely measuring several physical observables, is to extract entanglement measures from the quantum state. Key entanglement measures include the von Neumann entanglement entropy,
\begin{eqnarray}
S_{\rm vN}(\rho_A)=-{\rm Tr}\left(\rho_A\log \rho_A \right),\label{eq_vN}
\end{eqnarray}
and the R\'enyi entanglement entropy (indexed by $\alpha$),
\begin{eqnarray}
S_\alpha(\rho_A)=\frac{1}{1-\alpha}\log{\rm Tr}\left(\rho_A^\alpha\right),\label{eq_Ren}
\end{eqnarray}
for the reduced density matrix $\rho_A\equiv {\rm Tr}_{\bar{A}}(\rho)$ of subsystem $A$, when the whole system is in a pure quantum state. Here, ${\rm Tr}_{\bar{A}}(\cdot)$ denotes the partial trace with respect to the complement of $A$. The experimental detection of the entanglement entropy is of paramount importance, as it provides a versatile and precise measure of quantum correlations across various physical contexts. In condensed-matter physics, it is instrumental in identifying quantum critical points~\cite{Osterloh2002-ng,Eisert2010-ko} and topological order~\cite{Kitaev2006-vm,Wen2017-ax}. In quantum field theory, it reveals underlying structural properties, including the central charge and operator content, through its relationship with the entanglement spectrum~\cite{Calabrese2004-br,Lauchli2013-wz}. Moreover, entanglement entropy is deeply connected to black hole entropy in gravitational theory~\cite{Kabat1995-bs,Solodukhin2011-rv} and serves as a crucial indicator of information spread and thermalization processes in quantum thermodynamics~\cite{Kaufman2016-db,DAlessio2016-zf}.

Despite their critical importance, the experimental determination of entanglement measures remains challenging, particularly in large-scale systems. Direct measurements of entanglement have been rare, especially in more extensive systems. For instance, in cold-atom systems, while the second R\'enyi entanglement entropy $S_2(\rho_A)$ has been measured by letting two identical copies of the quantum state in a Bose–Hubbard system interfere, such measurements have so far been implemented in systems with a modest number of sites~\cite{Islam2015-rn,Kaufman2016-db} {(for  an experimental measure of $S_2(\rho_A)$ in trapped-ion systems see Ref.~\cite{brydges-19})}.

\begin{figure}[tb]
\begin{center}
\includegraphics[scale=0.42]{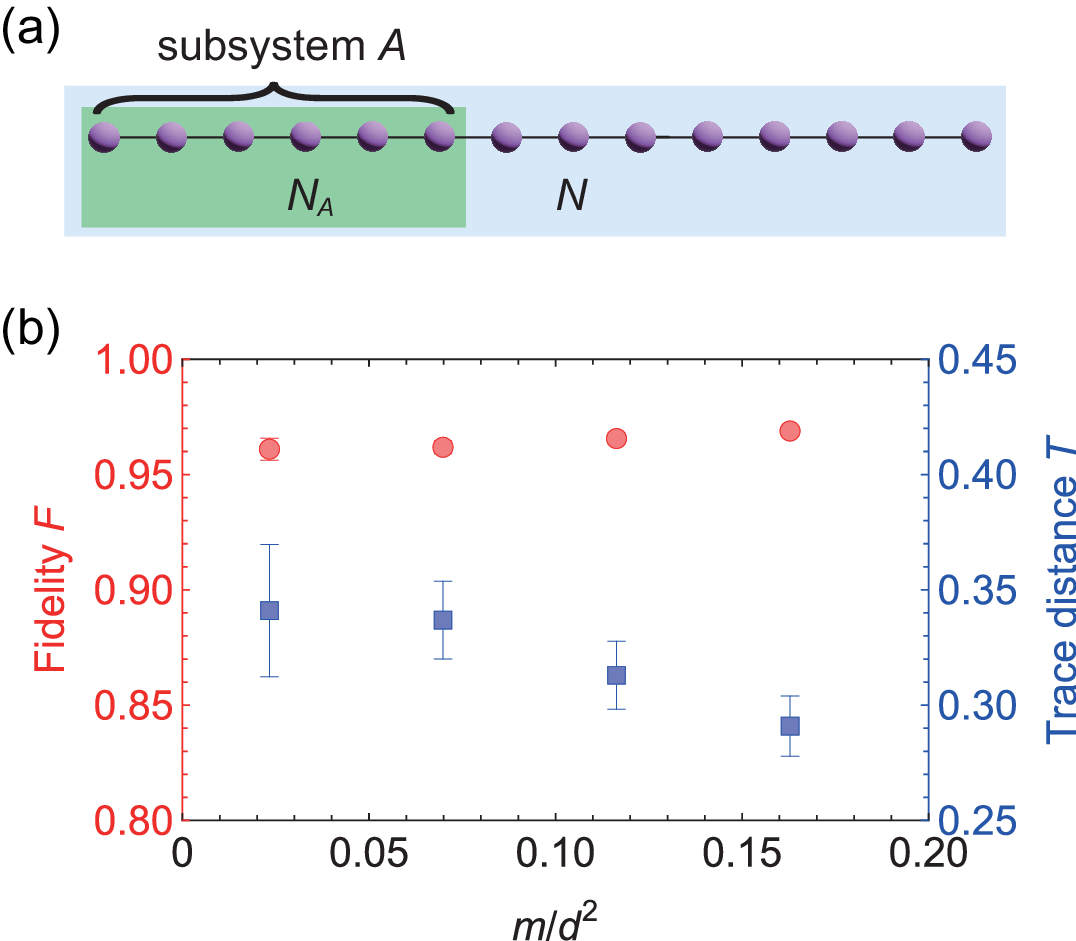}
\end{center}
\caption{{Spiral QST for the reduced density matrix.}  {(a)} Illustration of the bipartition of a system with $N$ sites into subsystem $A$, consisting of $N_A$ sites, and its complement $\bar{A}$. {(b)} Sample-averaged fidelity and trace distance for spiral compressed-sensing QST of the reduced density matrix $\rho_A={\rm Tr}_{\bar{A}}(\rho)$, where $N_A=7$, plotted as functions of the number of measurements $m$ scaled by $d^2=4^7$. The full density matrix $\rho$ corresponds to the ground state of the 14-site Heisenberg chain with NN antiferromagnetic spin-exchange coupling. {The number of repetitions for evaluating each expectation value is set to ${\rm reps}=500$}. The data points with the fewest measurements correspond to the spiral QST with  {$\{l=0\}$, followed by $\{l = 0, 1,-1\}$, $\{l = 0, 1,-1, 2,-2\}$,} and so on. }
\label{fig_reduced}
\end{figure}
We now demonstrate how entanglement measures in a quantum state can be extracted using spiral QST, which has the potential to overcome the challenge. Notably, to obtain the reduced density matrix $\rho_A$ via QST, it is not necessary to first reconstruct the entire pure state and then take a partial trace. Instead, the QST process can be applied directly to the measurement outcomes from the targeted subsystem $A$. To illustrate, we perform numerical simulations as follows: We consider a spin chain of length $N=14$, assumed to be prepared in the ground state $|\Psi_0\rangle$ of the Heisenberg model~\eqref{Heisenberg}, discussed in the previous subsection. Focusing on a subsystem $A$ consisting of sites from $i=1$ to $i=N_A$ [see Fig.~\ref{fig_reduced}{(a)}], we simulate the reconstruction of the reduced density matrix $\rho_A^{\rm rec}$ on $A$ via the spiral compressed-sensing QST. The von Neumann and R\'enyi entanglement entropies are then calculated from the reconstructed $\rho_A^{\rm rec}$, and compared with their theoretical values. 

A priori, it is not clear whether spiral QST (or, more generally, compressed sensing) is suitable for a reduced density matrix, as it typically represents a mixed state {of generic rank}. Therefore, we begin by comparing $\rho_A^{\rm rec}$ with the theoretical reduced density matrix $\rho_A$, which is obtained by tracing out the complement of $A$ from $\rho=|\Psi_0\rangle\langle\Psi_0|$. In Fig.~\ref{fig_reduced}{(b)}, we present the fidelity and trace distance between these matrices, with the subsystem size set to $N_A=7$ (half of the entire system). We {set} the number of repetitions for expectation value calculations {to} ${\rm reps}=500$. {The value of $\sigma_{\rm zp}$ does not affect the results because the reduced density matrix retains the SU(2) symmetry of the ground state $|\Psi_0\rangle$ of the entire system}. The results demonstrate that spiral QST is indeed effective also for the reduced density matrix, exhibiting a similar behavior to the case shown in Fig.~\ref{fig_Heisenberg}. This can be attributed to the strong hierarchy of eigenvalues, where a few dominant ones significantly outweigh the others, a characteristic commonly observed in physical systems, which renders this reduced density matrix effectively low-rank.

\begin{figure}[tb]
\begin{center}
\includegraphics[scale=0.42]{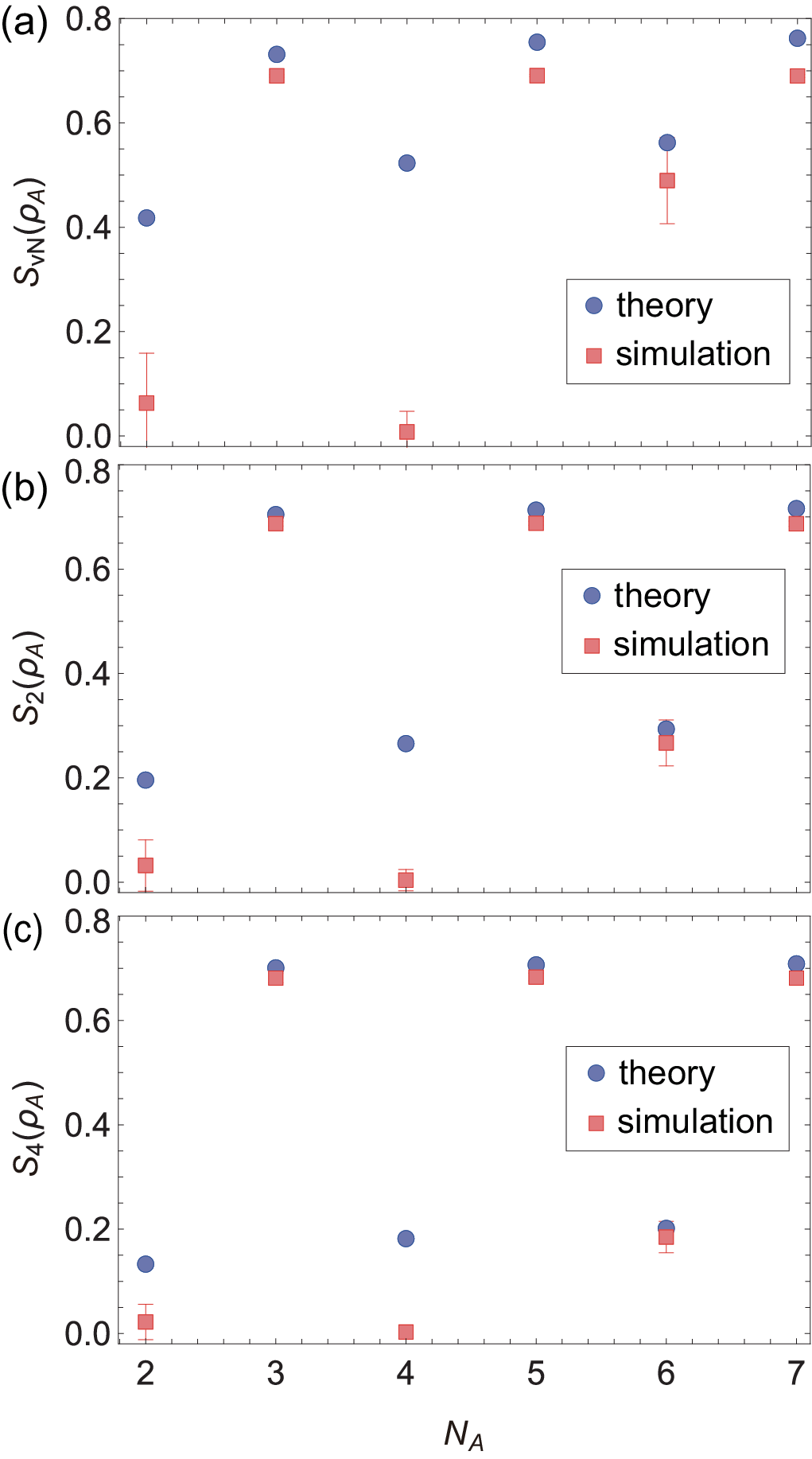}
\end{center}
\caption{{Entanglement entropies extracted by spiral QST for the NN Heisenberg chain.} ({a}) von Neumann entropy $S_{\rm vN}(\rho_A)$, ({b}) the second R'enyi entropy $S_2(\rho_A)$, and ({\bf c}) the fourth R'enyi entropy $S_4(\rho_A)$ of the reduced density matrix $\rho_A$ for the ground state of the 14-site Heisenberg chain with NN antiferromagnetic spin-exchange coupling, plotted as functions of the subsystem size $N_A$. We compare the values extracted from the reconstructed $\rho_A^{\rm rec}$ via spiral QST with $l=0$ to the theoretical  reference values. The amplitude of magnetic field fluctuations and the number of repetitions for evaluating each expectation value are fixed at $\sigma_{\rm zp}=0.1$ and ${\rm reps}=500$, respectively.}
\label{fig_EE}
\end{figure}
Now let us move on to the entanglement measures given in Eqs.\eqref{eq_vN} and~\eqref{eq_Ren}. In Figs.~\ref{fig_EE}{(a-c)}, we present the sublattice-size $N_A$ dependence of $S_{\rm vN}(\rho_A)$, $S_2(\rho_A)$, and $S_4(\rho_A)$, respectively, obtained from the reconstructed reduced density matrix $\rho_A^{\rm rec}$, {for the case of ${\rm reps}=500$ (for concreteness $\sigma_{\rm zp}=0.1$ is chosen)}, alongside the theoretical values. {Note that these are just representatives of the available entanglement measures; in fact, all the entanglement spectrum (and functions thereof) can be obtained from $\rho_A^{\rm rec}$ by simple diagonalization.} The target state {describes} a subsystem of size $N_A$ of the 14-site NN antiferromagnetic Heisenberg chain ($J_1=1$ and $J_{p\neq 1}=0$) in its ground state. For the reconstruction, we used only the measurements with $l=0$ ($m=3\times 2^{N_A} -2$), as the fidelity and trace distance are already sufficiently high and low, respectively, as seen in Fig.~\ref{fig_reduced}{(b)}.

First, we notice that the agreement between the values obtained from the reconstructed reduced density matrix and the theoretical ones is excellent for odd values of $N_A$, demonstrating the effectiveness of our spiral QST scheme in extracting entanglement properties of quantum states. However, for even values of $N_A$, the reconstructed $S_{\rm vN}(\rho_A)$ and $S_\alpha(\rho_A)$ are significantly underestimated. Notably, for $N_A = 2$ and $4$, it is incorrectly evaluated as $S_{\rm vN}, S_\alpha(\rho_A)\approx 0$, which would imply that subsystems $A$ and $\bar{A}$ are separable. 
This discrepancy is a consequence of the inherently weak entanglement when the system is partitioned into subsystems with even numbers of sites, a feature linked to the properties of the finite-size antiferromagnetic ground state considered here. Importantly, this issue does not arise from the spiral measurements but rather from the compressed sensing approach via the SVT method. Compressed sensing tends to lose information about the smaller eigenvalues of the reduced density matrix. For even $N_A$, in this case, the eigenvalues of the reduced density matrix comprise a dominant one  and the rest, which are at least one order of magnitude smaller. The compressed sensing procedure tends to erase these smaller contributions, almost ``purifying" the reduced state in the subsystem and thereby misrepresenting the true entanglement structure. %This explains why the simulated results tend to exaggerate the even-odd effect. This behavior is somewhat mitigated for the R\'enyi entanglement entropies $S_2$ and $S_4$ shown in Figs.~\ref{fig_EE}{\bf b} and~\ref{fig_EE}{\bf c}, as their dependence on the eigenvalues is less sensitive compared to $S_{\rm vN}$, where the logarithmic weighting amplifies the contribution of small eigenvalues.

Although the compressed sensing method has the aforementioned shortcoming, it may not pose a widespread practical problem. This issue, in fact, arises only in cases of very small entanglement between two subsystems, and such situations can be identified from the extracted entanglement entropies themselves.

\begin{figure}[tb]
\begin{center}
\includegraphics[scale=0.42]{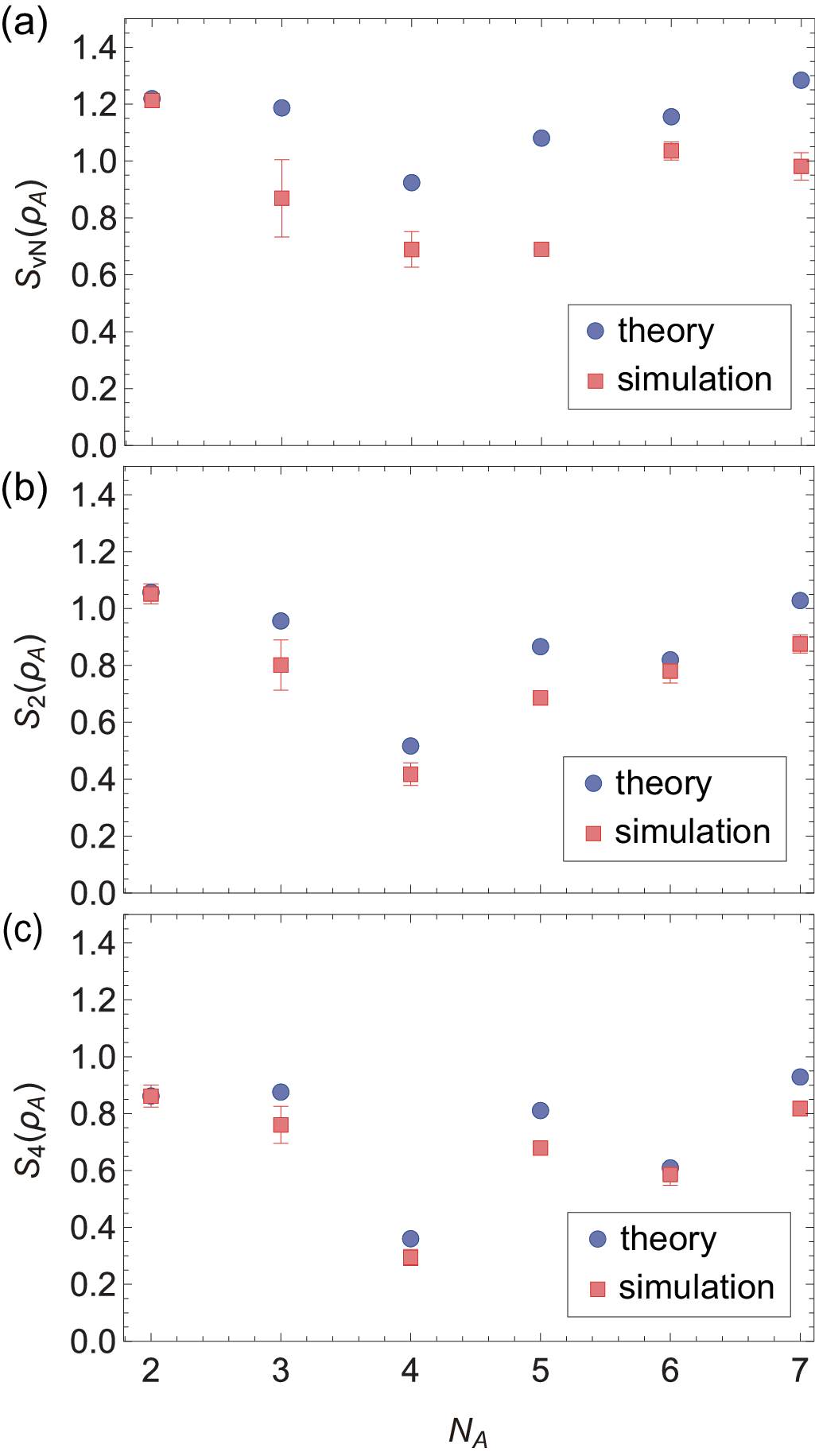}
\end{center}
\caption{{Entanglement entropies extracted by spiral QST for the frustrated NN+NNN Heisenberg chain.} ({a}) von Neumann entropy $S_{\rm vN}(\rho_A)$, ({b}) the second R'enyi entropy $S_2(\rho_A)$, and ({c}) the fourth R'enyi entropy $S_4(\rho_A)$ of the reduced density matrix $\rho_A$ for the ground state of the 14-site Heisenberg chain with frustrated NN and NNN antiferromagnetic spin-exchange coupling, plotted as functions of the subsystem size $N_A$. We compare the values extracted from the reconstructed $\rho_A^{\rm rec}$ via spiral QST with $l=0$ to the theoretical reference values. The amplitude of magnetic field fluctuations and the number of repetitions for evaluating each expectation value are fixed at $\sigma_{\rm zp}=0.1$ and ${\rm reps}=500$, respectively.}
\label{fig_EEJ1J2}
\end{figure}
Moving beyond the previous simple model, we now demonstrate a more interesting case with antiferromagnetic NN coupling $J_1>0$ and next-nearest-neighbor (NNN) coupling $J_2>0$. These interactions compete with each other, leading to frustrations among the spins, which is expected to induce a larger entanglement in the ground state. In experiments with optical lattices, such a model can be realized by arranging the laser beams in a way that creates a zig-zag chain configuration~\cite{Zhang2015-bh}. Note that the zig-zag structure does not interfere with the procedure involving a magnetic field gradient along the chain direction. 

We present the results for $J_1=J_2=1$ and $J_{p\neq 1,2}=0$ in Figs.~\ref{fig_EEJ1J2}(a-c). 
%Although the von Neumann entropy still proves to be {relatively} challenging due to the logarithmic weighting that amplifies the errors, 
These plots reveal how well the spiral QST {can capture} the entanglement properties of the ground state of quantum many-body systems. 
{Furthermore, the fact that the relative errors get better for higher Renyi index reinforces the notion that the reconstruction misses out only on the tail of small eigenvalues.}
These demonstrations provide valuable insights into the effectiveness of the method in important and interesting scenarios, such as frustrated quantum systems. {To reinforce the efficiency of spiral QST in estimating this kind of nonlinear functions of the (reduced) density matrix, let us also note that the relative error achieved in Fig.~\ref{fig_EEJ1J2}, together with the low total number of measurements (including repetitions) used, is in line with the best results available in the literature for target states with similar amounts of entanglement (see in particular Fig.~4b in Ref.~\cite{huang-20}).}

\section{Conclusions}
We have proposed and {validated} an efficient method for quantum state reconstruction using compressed sensing based on spiral measurements. This method does not require individual single-qubit addressing, yet our numerical simulations on random target states showed accuracy comparable to that of traditional Pauli measurements. A key advantage of our spiral QST approach is its scalability: different measurement setups with varying spiral pitch angles can be achieved simply by adjusting the application time or magnetic field strength using a single, globally acting magnetic field generator. Unlike conventional Pauli measurements, which require increasingly complex one-by-one qubit addressing as the system size grows, spiral QST remains efficient and practical as quantum platforms continue to scale. This scalability advantage becomes particularly significant as quantum systems expand in size and complexity. Our method is particularly effective for systems where single-qubit addressing is challenging and prohibitively expensive, making it a versatile tool for a broad range of quantum systems. {Also,  while the localized spin limit is assumed so far, it is conceivable that the method might be extendable to states with motional degrees of freedom, such as holes and doublons in Hubbard and $t$-$J$ models, which are subjects of intense study in optical-lattice quantum simulators \cite{prichard-24,lebrat-24,bourgund-25,xu-25}.}

In addition,
%to its scalability, 
the spiral QST method has proven robust against realistic noise sources. When applied to the ground state of specific Hamiltonians, such as the Heisenberg model, it maintains its effectiveness even in the presence of characteristic experimental inaccuracies. For instance, we have considered scenarios with realistic magnetic field fluctuations and a finite number of measurement repetitions, demonstrating that spiral QST can still faithfully reconstruct the quantum state. This robustness is crucial for practical implementations, where noise and errors are inevitable. Furthermore, the spiral QST approach is suitable for extracting entanglement measures, a key objective of QST in quantum simulations of various fields of physics. The accurate determination of entanglement measures such as von Neumann and R\'enyi entropies is {critical}, as these quantities provide deep insights into the quantum correlations and properties of the system. 

{Concerning concrete  future directions, it would be interesting to look  at the dynamics of entanglement and correlations in various physical situations (quantum quenches, local excitations of quasiparticles, etc.), in particular focusing on the relation between the intrinsic timescales of the target system and those of the measurement process. For example, one can imagine devising an experimental protocol to observe the quantum Mpemba effect in an optical-lattice setup, via the estimation of entanglement asymmetry obtained from the spiral QST of the reduced density matrices \cite{ares-23}. Also, in view of the recent interest for measurement-induced
quantum phase transitions, originally formulated within gate-based quantum circuits (for which spiral QST might be a useful tool depending on the platform),  one may think of conceptually related phenomena in optical lattices, where the unitary evolution is given by a Hamiltonian and the non-unitary projection by particle losses by inelastic collisions \cite{syassen-08}, electron beams \cite{barontini-13}, photoassociation \cite{tomita-17}, etc. Spiral QST is suited to approach the problem from the area-law side, and could possibly access the shallow volume-law region when the system size is relatively small; it should be interesting to see the dependence of the entanglement entropy (or entanglement negativity for mixed states) on the system parameters, {\it e.g.} the dissipation rates.}

Despite significant advances, quantum processors are still in a developmental phase, with continuous improvements in error correction and fault tolerance as they progress toward fully scalable and general-purpose quantum computing~\cite{RevModPhys.95.045005}. During this period of transition, quantum processors are proving their remarkable ability to simulate complex physical systems, demonstrating their practical utility and potential~\cite{Houck2012-hc,Daley2022-ok,Kim2023-nz,Frey2022-mp,Mi2022-pl,Google_Quantum_AI_and_Collaborators2023-bc,Weimer2010-fu,Henriet2020-yk,Gross2017-rb,Barreiro2011-cp,Blatt2012-fa,Monroe2021-tn}. The present work not only illustrates the practical application of spiral QST, but also provides a framework for overcoming the experimental difficulties associated with direct measurements of entanglement in large-scale systems. This capability will contribute to a comprehensive understanding of quantum systems and foster interdisciplinary research by bridging condensed matter physics, quantum field theory, gravitational theory, quantum thermodynamics, and other fields through the lens of quantum information science.

\begin{acknowledgments}
We would like to thank S.~Minagawa, H.~Ozawa, Y.~Kakihara, and K.~Baba for useful discussions. The work of G.\ M.\ was supported by JSPS KAKENHI Grant No.~21H05185. The work of T.\ F.\ was supported by JSPS KAKENHI Grant No.~23K25830, JST ERATO Grant No. JPMJER2302, and JST Moonshot R\&D Program Grant No. JPMJMS2269. The work of D.\ Y.\ was supported by JSPS KAKENHI Grant Nos.~21H05185, 23K22442, 23K25830, 24K06890, and JST PRESTO Grant Nos.~JPMJPR2118 and JPMJPR245D. 
\end{acknowledgments}

\section{Author contributions}
D.Y. designed and coordinated the project. G.M. developed the applications to entanglement measures, with additional input from D.Y. G.M. devised the algorithms, performed numerical simulations and analyzed the data. T.F. provided insights mainly from an experimental perspective. D.Y. primarily wrote the manuscript, with additional input from G.M. and T.F. All authors reviewed and approved the final version of the manuscript.

\appendix

\section{Singular value thresholding}
\label{appendixA}
The quantum compressed sensing protocol \cite{gross-11} is aimed at reconstructing a target low-rank density matrix $\rho$ from an under-complete set of tomographic data, namely  a set of expectation values of linearly independent operators (measurement operators). Mathematically, the problem is formulated in terms of a  matrix variable $\sigma$  as the minimization of {the trace norm} $||\sigma||_1$, subject to ${\rm Tr} (\sigma)=1$ and ${\rm Tr} (\sigma w_a)= {\rm Tr} (\rho w_a)$, $a=1,\ldots,m$, where $\{w_a\}$ are the measurement operators used. We denote the outcome of this problem as $\rho^{\mathrm{rec}}$, the reconstructed density matrix.

In practical numerical calculations, one employs the singular value thresholding (SVT) algorithm \cite{Cai2008-zn} to minimize $\tau ||\sigma||_1 +||\sigma||^2_2/2 ${, where $||A||_2\equiv \sqrt{{\rm Tr}(A^\dagger A})$ is the Frobenius norm,} subject to the above constraints. Clearly this reduces exactly to the minimization of the trace norm for $\tau\to \infty$, but in actual implementations it is enough to take a sufficiently large, but finite, $\tau$; throughout this work $\tau=5$. 

Target quantum states are generated as follows. Pure states of dimension $d$ are simply obtained by applying a random unitary matrix drawn from the Haar measure on SU$(d)$ to a fixed reference state. Mixed states of rank $r$ are obtained by first generating a Haar-random pure state in $d 
\times r$ dimensions and then partial-tracing over the auxiliary $r$ degrees of freedom; this induces a natural measure on the space of rank-$r$ states \cite{zyczkowski-01,bengtsson-06,miszczak-12}. An efficient way to reach the same result is generating a random $d \times r$ matrix $G$ from the Ginibre ensemble \footnote{The Ginibre ensemble is constituted by random complex matrices whose entries (both real and imaginary parts) are i.i.d. variables distributed according to the standard normal distribution $\mathcal{N}(0,1)$.} and then defining the target state as $\rho=G G^\dagger /{\rm Tr}(G G^\dagger)$. Finally, the ground states of the Hamiltonians described in the text are calculated by exact diagonalization; the reduced density matrices are then computed via partial tracing over the complement of the subsystem of interest. We have included a component of depolarizing noise to the random states as described in Sec.~\ref{sec2} 
to mimic the inaccuracies that may normally occur in the preparation of an actual quantum state; we have not done so for the Hamiltonian ground states and related reduced density matrices,
%Sec.~\ref{sec:simexp}-\ref{sec:detee}, 
since we intend to focus on the fluctuations of the measurement processes involving application of magnetic field gradients and QGM.

Our SVT codes are tailored to the use of the spiral operators $\{\tilde{M}^{\alpha\beta}_{\bm{a}}(q)\}$ as measurement operators. 
Let us mention that this set is not orthogonal, but ``almost'' orthogonal, in the sense that the number of non-orthogonal pairs scales like $1/2^N$; it is however linearly independent, except for a few accidental duplications (up to a sign), {\it e.g} $\tilde{M}^{XY}(0)=X \otimes \cdots \otimes X = \pm \tilde{M}^{ZX}(\pi)$. These duplications, when they occur, are taken care of before the actual SVT calculation.
In each simulation, a certain number $n_q$ of pitch angles is chosen from the set in Eq.~\eqref{pitch}. This choice is random for the analysis of random states, whereas for the Hamiltonian ground states and related reduced density matrices we adopt the strategy of relevant pitch angles described in the following subsection. For each chosen $q$ we use all the {non-trivial} $2^N-1$ spiral operators in all of the 3 spin planes.  Note that our sets of measurements are therefore very structured, as opposed to the completely random sets pertaining to the original quantum compressed sensing algorithm \cite{gross-11}. 

The expectation values used in the QST of random quantum states are the subset of $\{{\rm Tr}(\rho \tilde{M}^{\alpha\beta}_{\bm{a}}(q))\}$ dictated by the choice of $q$'s, with the addition of Gaussian fluctuations of width $\sigma=0.1/d$, as described in Sec.~\ref{sec2}. In the simulation of the optical-lattice setting with QGM, the derivation of the input expectation values deserves a few more words. Upon the repeated preparation of a quantum state, a  series of QGM snapshots naturally leads to the measurement of correlations in the $Z$ basis, that is expectation values of the operators $\tilde{M}^{ZX}_{\bm{a}}(0)$. Let us consider that, given any family of operators $\tilde{M}^{\alpha\beta}_{\bm{a}}(q)$,
%associated with a different spin plane and/or $q$, 
there is a well-defined unitary transformation $U=U_1\otimes\cdots\otimes U_N$ such that  $\tilde{M}^{\alpha\beta}_{\bm{a}}(q) = U \tilde{M}^{ZX}_{\bm{a}}(0) U^\dagger$ ($U$ depends on $\alpha\beta$ and $q$, but we omit these indices for simplicity of notation). It follows that
\begin{eqnarray}
 &  {\rm Tr}(\rho\, \tilde{M}^{\alpha\beta}_{\bm{a}}(q)) & =  {\rm Tr}(\rho\, U \tilde{M}^{ZX}_{\bm{a}}(0) U^\dagger) \nonumber\\
&&    =     {\rm Tr}(U^\dagger \rho\, U \tilde{M}^{ZX}_{\bm{a}}(0)).
\end{eqnarray}
This means that the new expectation values can be obtained by operating on the target state with $U^\dagger$, namely $U^\dagger \rho U \equiv \rho^\prime$. This is exactly what the optical (Rabi pulses) and magnetic (field gradient pulses) operations achieve. We include the Gaussian fluctuations adopted for the zero point of the magnetic field (see Sec.~\ref{sec:realworld}) into all our numerical implementations of $U$. At this stage,  the diagonal elements of $\rho^\prime$ simply give a probability distribution for the QGM snapshots, {\it e.g.} $\rho^\prime_{11}$ is the probability of obtaining $|\uparrow \cdots \uparrow\rangle$,  $\rho^\prime_{22}$ is the probability of obtaining $|\uparrow \cdots \uparrow \downarrow\rangle$, etc. In the numerical simulations, we draw a sufficient number (denoted ``reps'') of snapshots from this distribution to evaluate each family of expectation values $\langle \tilde{M}^{\alpha\beta}_{\bm{a}}(q)\rangle$. {In statistics and probability theory, ``reps'' is basically the sample complexity of learning the distribution; for a fixed distance to the true distribution (measured by the $l_1$ distance, the classical equivalent of the trace distance), it is expected to scale linearly with $d$ \cite{canonne-20}. This is compensated by the fact that the number of obtained independent expectation values also has the same scaling.}

%\begin{widetext}
\begin{table*}[htb]
    \centering
    \begin{tabular}{|c|c | c |c|}
    \cline{2-4}
     \multicolumn{1}{ c | }{\it Target states / SVT parameters} & $\delta$ & $k_{max}$ & $s$ \\
     \hline
    Random pure (Fig.~\ref{fig_random}{(a)})     & 1.5  & 100 & 25 \\
    \hline
   \multirow{3}{*}{Random rank=3  (Fig.~\ref{fig_random}{(b)})}  & 4 for $n_q\geq 12$  & \multirow{3}{*}{100} & \multirow{3}{*}{25} \\
     & $48/n_q$ for $9\leq n_q \leq 11$ & & \\
    & $0.4(n_q-8)+5$ for $n_q\leq 8$ & & \\
     \hline
    $N=8$ NN Heisenberg g.s. (Fig.~\ref{fig_Heisenberg}) & 0.3(-0.1 $n_q$ +1.1) & 100 & 10 \\
    %\hline
    %$N=6$ NN Heisenberg+DM g.s. (Fig.~\ref{fig_DM}) & 1/$n_q$ & 150 & 10\\
     \hline
    \multirow{2}{*}{$N=8$ NN Heisenberg+DM g.s. (Fig.~\ref{fig_DM}) } & -0.1 $n_q$ +1.1 for $\sigma_{zp}=0.01,0.1$ & \multirow{2}{*}{100} & \multirow{2}{*}{10} \\
     & -0.07 $n_q$ +1.07 for $\sigma_{zp}=0.5,1.0$ & & \\
    \hline
    \multirow{6}{*}{NN Heisenberg, reduced (Figs.~\ref{fig_reduced} and \ref{fig_EE}) } & $1/n_q$ for $N_A=2$ & \multirow{6}{*}{100} & \multirow{6}{*}{50} \\
     & $0.5/n_q$ for $N_A=3$ & & \\
      & $0.7/n_q$ for$N_A=4$ & & \\
       & $1.3/n_q$ for $N_A=5$ & & \\
        & $2.2/n_q$ for $N_A=6$ & & \\
         & ${1.5}/n_q$ for $N_A=7$ & & \\
    \hline
    \multirow{6}{*}{NN+NNN Heisenberg, reduced (Fig.~\ref{fig_EEJ1J2}) } & $1/n_q$ for $N_A=2$ & \multirow{6}{*}{100} & \multirow{6}{*}{10} \\
     & $1.2/n_q$ for $N_A=3$ & & \\
      & $1.2/n_q$ for $N_A=4$ & & \\
       & $0.8/n_q$ for $N_A=5$ & & \\
        & $1.2/n_q$ for$N_A=6$ & & \\
         & $2.2/n_q$ for $N_A=7$ & & \\
    \hline
    \end{tabular}
    \caption{Optimized SVT parameters $\delta,k_{max}$ for the various quantum states of interest in this work. $s$ is the  number of simulations per data point in the corresponding figures illustrating spiral QST results and related entanglement measures.}
    \label{tab:svtpar}
\end{table*}
%\end{widetext}

For the details of the SVT routine itself we refer to the original paper \cite{Cai2008-zn}. Let us just recall that it is an iterative procedure and its convergence, at least in our applications, is extremely sensitive to the choice of an internal parameter, denoted $\delta$, which can be seen as the ``size'' of an iteration step. The fine-tuning of $\delta$ as a function of system size, target states and $n_q$ is crucial in order to obtain a good $\rho^{\mathrm{rec}}$; our best choices of $\delta$ for each case study are reported in Table~\ref{tab:svtpar}. {A preliminary indicator of the goodness of $\delta$ is the trace of the reconstructed matrix being close to one (before the final normalization) and having a limited spread around the mean value over various simulations.} The iteration can terminate either if a control condition is met or a maximum number of steps $k_{max}$ is reached (see Table~\ref{tab:svtpar}). The control condition is $(\sum_{a=1}^m ({\rm Tr} ((\rho^{\mathrm{rec}} -\rho) w_a))^2 )^{1/2}<\epsilon$; we take $\epsilon=10^{-1}$. The output $\rho^{\mathrm{rec}}$ in general does not have unit trace (it is typically smaller) and therefore
we normalize it {\it a posteriori} to have a {\it bona fide} density matrix, of which it is possible to calculate fidelity and trace distance with the target state $\rho$. Upon repeating the whole process a number  $s$  of times (see Table~\ref{tab:svtpar}), one can extract a data point and its error bar in Figs.~\ref{fig_random}-\ref{fig_reduced}, corresponding to the mean value and standard deviations of the sampled fidelity and trace distance. As for Figs.~\ref{fig_EE}-\ref{fig_EEJ1J2}, a sample of $\rho^{\mathrm{rec}}$'s is produced in a similar way and their entanglement entropies are statistically analyzed.

\section{Strategy of relevant pitch angles}
\label{appendixB}
The strategy of relevant pitch angles consists in selecting a subset of pitch angles that are most likely to contribute significantly to the quantum state's characterization, based on the system's symmetry and the expected distribution of ``tomographic weight'' among the $q$'s in Eq.~\eqref{pitch}. By focusing on these relevant pitch angles, the spiral QST process becomes more efficient, as it prioritizes measurements that are more likely to yield important information about the quantum state. This approach reduces the number of necessary measurement setups, while still capturing the essential features of the state, especially for quantum states that exhibit a strong hierarchy in the weight of the various $q$'s or certain symmetries that restrict the range of relevant pitch angles.

\begin{figure}[bt]
\begin{center}
\includegraphics[scale=0.42]{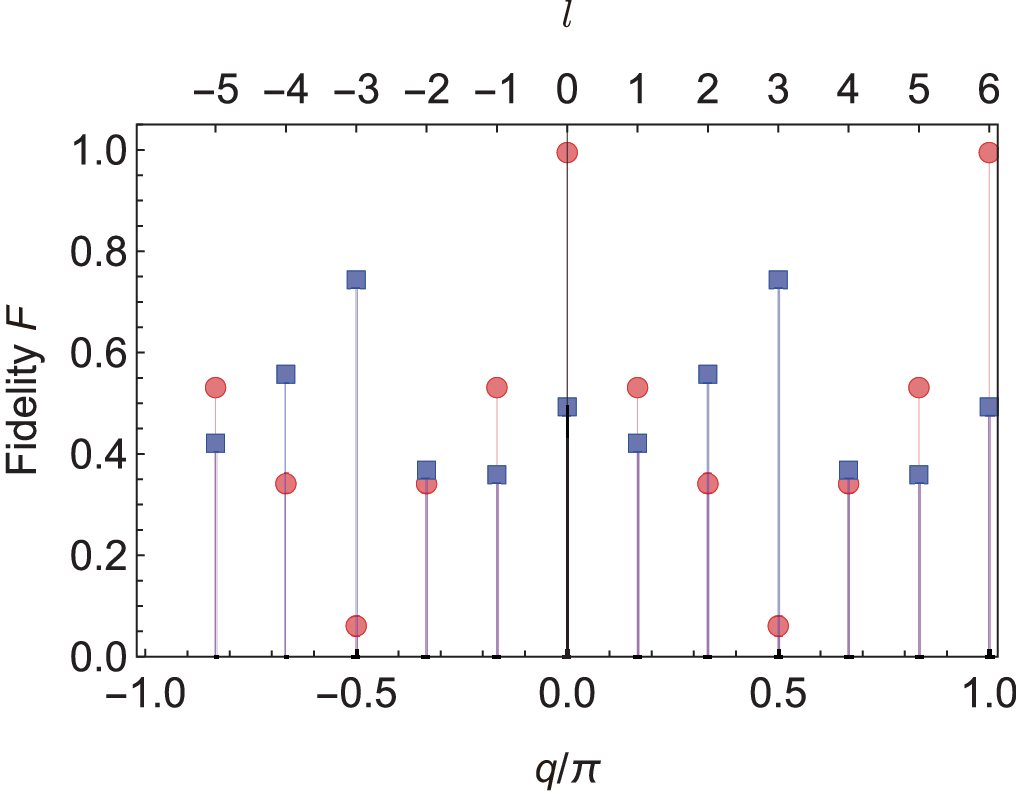}
\end{center}
\caption{{QST weight of individual pitch angles.} Fidelity for the state tomography of the ground state of typical spin Hamiltonians  with only a single pitch angle $q=\pi l/N$ ($l=-N+1, -N+2, \dots, N$), under ideal state preparation and spiral measurements conditions. Results are shown for the $N=6$ antiferromagnetic Heisenberg chain with $J_1 = 1$, $J_{p \neq 1} = 0$, $D = 0$ (circle symbols) and the $N=6$ ferromagnetic Heisenberg + DM chain with $D = -J_1 = 1$, $J_{p \neq 1} = 0$ (square symbols).}
\label{fig_pitch}
\end{figure}

In a practical approach, one initially performs simulations on smaller systems  using classical computers to identify which pitch angles are most significant. Once these key parameters are determined, they can be effectively applied to experiments on larger systems. This strategy ensures that the most relevant aspects of the quantum state are accurately captured in larger-scale implementations.

As examples, in Fig.~\ref{fig_pitch}, we show the QST weight for each pitch angle $q$ for the ground states of the antiferromagnetic Heisenberg chain and the ferromagnetic Heisenberg + DM chain with $N=6$ sites, assuming ideal state preparation and measurements. The QST weight is evaluated using the fidelity between the exact density matrix obtained by the exact diagonalization and the one reconstructed with only a single pitch angle $q$, selected from Eq.~\eqref{pitch}. 

For the Heisenberg chain, the spin symmetry of the system leads to additional periodicity and reflection symmetries as a function of $q$. While this behavior is specific to this system, many important quantum states in typical quantum many-body systems exhibit similar symmetries. Based on this observation, it is reasonable to begin QST with {$\{l=0\}$, then expand sequentially to $\{l=0,1,-1\}$, $\{l=0,1,-1,2,-2\}$, and so on for such symmetric systems. It is interesting to note that, if we also include the sets $\{0,1\},\{0,1,-1,2\}$, etc., which do not respect the symmetry, we  encounter instances of non-monotonicity of fidelity and trace distance as a function of $m/d^2$. This indicates, to some extent unexpectedly, that the algorithm is indeed quite sensitive to whether or not the choice of $q$'s is symmetric. Incidentally, the above sequence of symmetric $q$-sets is used also for the reduced density matrix (see Fig.~\ref{fig_reduced}), which turns out to have an approximate inversion symmetry.}

In the presence of DM interactions, the tomographically dominant pitch angle shifts to a finite $q$. In this case (see Fig.~\ref{fig_pitch}'s caption for the value of the couplings), the contribution of $q = \pi/2$ (which coincides with that of $q = -\pi/2$ due to a specific symmetry) appears to be dominant. Consequently, starting spiral QST from pitch angles around $q = \pi/2$  represents a clever and efficient strategy in such cases, {but the question is how to add the subsequent pitch angles. One may choose to progressively add the adjacent ones in the following fashion: (for $N=8$) $\{l=4\}, \{4,5\}, \{4,5,3\},\{4,5,3,6\},\{4,5,3,6,2\}$, etc.. This is reminiscent of the sequence for the Heisenberg case, with $l=4$ in the role of $l=0$. Hover this leads to a tomographic sequence that is worse overall compared to that in Fig.~\ref{fig_DM} (except for the first data point, as argued above). One possible way to effectively improve the spiral QST for the DM case even further is to modify the algorithm so that the pitch angle of the measurement operators in the $XY$ (where the spin spiral characterizing the ground state develops) is independent of those in the $YZ$ and $ZX$ plane.}

%\section{Data availability}
%The data that support the findings of this study are available from the corresponding author upon reasonable request.

%\bibliographystyle{unsrt}
%\bibliographystyle{naturemag}

%\section{Competing interests}
%The authors declare no competing interests.

%\section*{Additional information}

\end{document}